\documentclass[aps,prb,reprint,showpacs]{revtex4-1}
\usepackage[dvipdfmx]{graphicx}
\usepackage{subfigmat}
\usepackage{amsmath,amssymb}
\usepackage{color}
\usepackage{bm}
\usepackage{hyperref}
\usepackage{comment}
\usepackage{braket}
\usepackage[normalem]{ulem}

\usepackage{upgreek}

\begin{document}

\title{Spin-Hall magnetoresistance in quasi-two-dimensional \\  antiferromagnetic insulator/metal bilayer systems}

\author{T. Ishikawa$^{1}$, M. Matsuo$^{2,3,4,5}$, and T. Kato$^{1}$}
\affiliation{
${^1}$Institute for Solid State Physics, The University of Tokyo, Kashiwa, Japan \\
${^2}$Kavli Institute for Theoretical Sciences, University of Chinese Academy of Sciences, Beijing, China \\
${^3}$CAS Center for Excellence in Topological Quantum Computation, University of Chinese Academy of Sciences, Beijing, China\\
${^4}$Advanced Science Research Center, Japan Atomic Energy Agency, Tokai, Japan\\
${^5}$RIKEN Center for Emergent Matter Science (CEMS), Wako, Saitama, Japan\\
}

\date{\today}

\begin{abstract}
We study the temperature dependence of spin Hall magnetoresistance (SMR) in antiferromagnetic insulator (AFI)/metal bilayer systems.
We calculate the amplitude of the SMR signal by using a quantum Monte Carlo simulation and examine how the SMR depends on the amplitude of the spin, thickness of the AFI layer, and randomness of the exchange interactions. 
Our results for simple quantum spin models provide a useful starting point for understanding SMR measurements on atomic layers of magnetic compounds.
\end{abstract}
\maketitle

\section{Introduction}
\label{sec:introduction}

In the research field of spintronics, various types of magnetoresistance, such as giant magnetoresistance~\cite{Baibich1988,Binasch1989,Fert2008} and tunneling magnetoresistance~\cite{Julliere1975,Miyazaki1995,Moodera1995,Yuasa2004,Parkin2004}, have been used in devices for sensors, memories, and data storage. Recently, a novel type of magnetoresistance, called spin Hall magnetoresistance (SMR), has been attracting much attention.
SMR was first observed in a normal metal (NM)/ferromagnetic insulator (FI) bilayer\cite{Nakayama2013a,Chen2013a,Hahn2013,Vlietstra2013,Althammer2013,Meyer2014,Marmion2014,Cho2015,Kim2016,Chen2016,Sterk2019,Tolle2019} and subsequently in a NM/antiferromagnetic insulator (AFI) bilayer\cite{Hou2017,Lin2017,Cheng2019,Hoogeboom2017a,Fischer2018,Ji2018,Lebrun2019}. Because SMR reflects information on the FI magnetization (or AFI N\'{e}el vector), it can be utilized for detecting the orientation of ordered spins in magnetic materials.

SMR also has the potential to be a useful probe for two-dimensional magnetic materials, such as van der Waals atomic layers.
In fact, several two-dimensional atomic layer compounds showing ferromagnetism\cite{2d_fi} and antiferromagnetism\cite{2d_afi} have recently been synthesized recently. Since SMR measurements on such magnetic atomic layers have been performed in recent experiments\cite{Wu2021,Feringa2022}, it has become an urgent task to construct a theory of SMR that is applicable to two-dimensional quantum magnets.

\begin{figure}[tb]
 \begin{center}
  \includegraphics[clip,width=7.0cm]{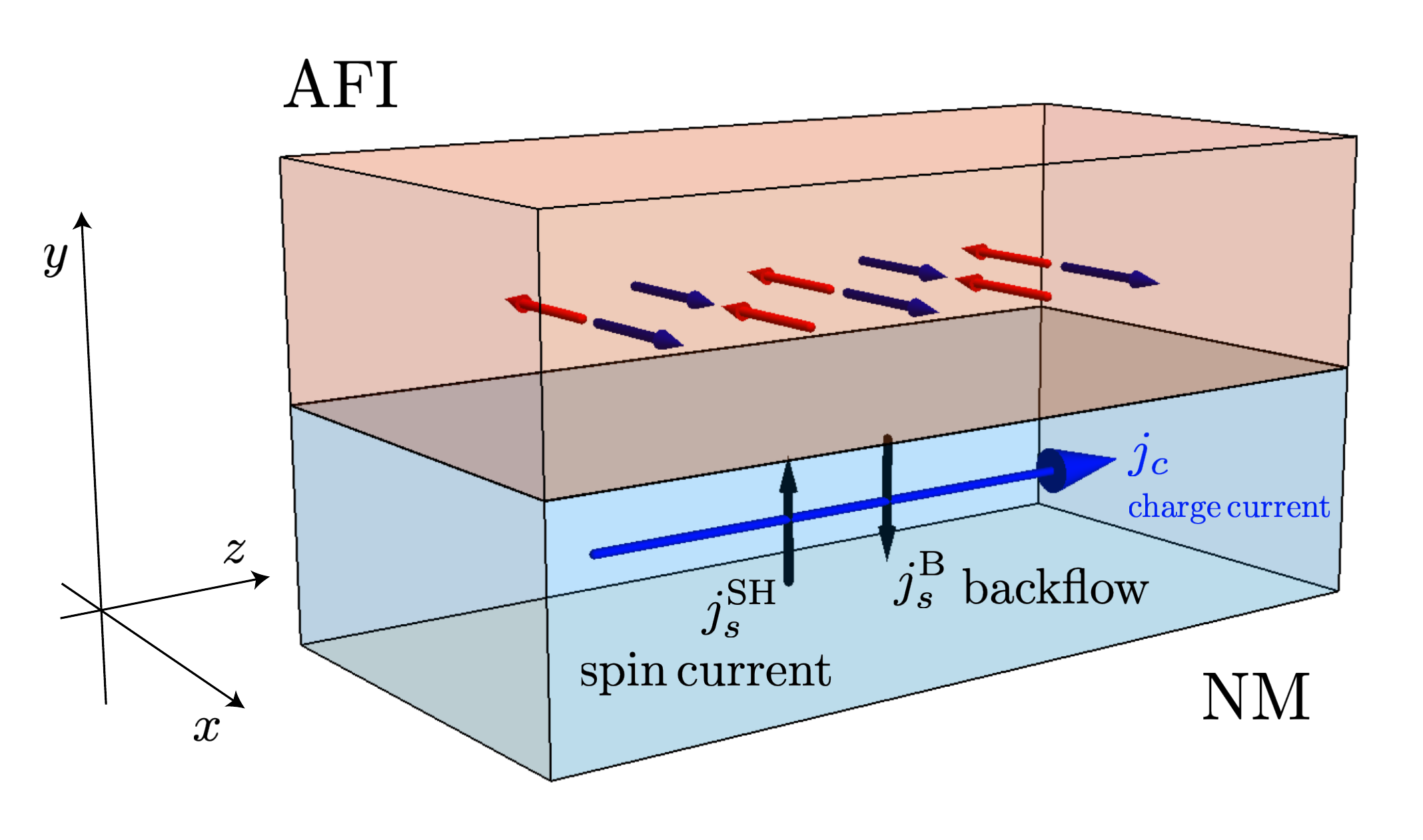}
    \caption{Illustration of spin Hall magnetoresistance. Spin absorption at the interface changes according to the orientation of the Neel vector of the antiferromagnetic insulator, and the magnitude of the magnetoresistance changes.}
    \label{fig:SMR_setup}
  \end{center}
\end{figure}

The existing theory\cite{Nakayama2013a,Chen2013a} explains SMR as follows (see Fig.~\ref{fig:SMR_setup}).
When an in-plane charge current flows in the NM layer, spin accumulation occurs near the NM/FI(AFI) interface due to the spin Hall effect\cite{Sinova2015}.
Then, a backflow spin current, which is induced by spin diffusion in the NM, is converted into the charge current again by the inverse spin Hall effect (ISHE), resulting in longitudinal magnetoresistance.
This longitudinal magnetoresistance depends on the orientation of the FI magnetization (or AFI N\'{e}el vector),
because the amount of spin that accumulates at the interface is changed by it.
Note that the strength of SMR is proportional to square of the spin Hall angle of the NM.

Although this semiclassical theory \cite{Nakayama2013a,Chen2013a} succeeds in explaining the qualitative features of SMR measurements, it does not explain the temperature dependence of the SMR signal.
Recently, two of the authors constructed a microscopic theory using Green's function~\cite{Kato2020}. 
This microscopic theory describes SMR in terms of local spin susceptibilities of the FI(AFI).
In particular, it can include dynamic processes such as magnon absorption and emission responsible for a nontrivial sign change in the SMR signal, which are neglected in the semiclassical theory.

In Ref.~\onlinecite{Kato2020}, SMR was calculated analytically by employing the spin-wave approximation.
However, this approximation cannot be applied near the transition temperature.
Furthermore, it becomes inaccurate when the magnitude of the localized spin in the FI(AFI) is small.
Therefore, to obtain the features of SMR in a wide range of the temperature and at an arbitrary magnitude of the localized spin, $S$, we need to calculate it without employing the spin-wave approximation.

In the study reported here, we numerically calculated SMR by using a quantum Monte Carlo method based on the formulation in Ref.~\onlinecite{Kato2020}.
We utilized a new method for accurately obtaining the integral of the local spin susceptibilities from the QMC data without using numerical analytic continuation.
We evaluated the detailed temperature dependence of SMR for $S=1/2$ and $S=1$ spin systems on a two-dimensional square lattice and a quasi-two-dimensional cubic lattice with a finite number of layers.
On the basis of the numerical results for these models, we examined the qualitative features of SMR.

This study is organized as follows. In Sec.\ref{sec:formulation}, we formulate SMR in terms of a microscopic theory based on a microscopic Hamiltonian of NM, AFI, and exchange interaction at the interface. In Sec. \ref{sec:Numerical}, we propose an accurate numerical calculation method for SMR. 
In Sec.\ref{sec:Result}, we show numerical results obtained from a quantum Monte Carlo simulation.
Finally, we discuss experimental relevance of our work in Sec.~\ref{sec:discussion} and summarize our results in Sec.~\ref{sec:summary}.
We provide the details of the derivation and a comparison with theoretical calculation in the Appendixes.

\section{Formulation}
\label{sec:formulation}

\subsection{Spin conductance}
\label{sec:SpinConductance}

First, let us summarize the theoretical framework for SMR~\cite{Chen2013a,Kato2020}, that is employed in this study.
When an electric field is applied to the metal side in the $x$ direction, a charge current $j_x$ induces a spin current $j_s^{\rm SH}=\theta_{\rm SH} j_x$ in the $y$ direction due to the spin Hall effect, where $\theta_{\rm SH}$ is the spin Hall angle. 
As a result, the electron spins parallel to the $z$-axis are accumulated near the interface.
This spin accumulation is described by the spin chemical potential $\mu_{s}(y)=\mu_{\uparrow}(y)-\mu_{\downarrow}(y)$, where $\mu_{\uparrow}(y)$ and $\mu_{\downarrow}(y)$ are the chemical potentials of two different spins. 
The spatial gradient of $\mu_{s}(y)$ generates a backflow $j_s^{\rm B}$ in the $y$ direction due to spin diffusion.
Finally, this backflow is converted into a charge current in the $x$ direction due to the inverse spin Hall effect, which results in SMR. 

The amount of spin that accumulates near the interface is affected by the spin loss rate at the interface $I_{S}$, which is assumed to be
proportional to the spin chemical potential $\mu_{s}(0)$, where the position of the interface is set as $y=0$.
Following Ref.~\onlinecite{Kato2020}, we introduce the spin conductance $G_s$ as
\begin{align}
G_s &= \lim_{\mu_{s}(0)\rightarrow 0} \frac{I_{S}}{\mu_{s}(0)}.
\label{eq:spinconductancedef}
\end{align}
Note that $G_s$ depends on the orientation of the N\'{e}el vector of the AFI.
From the above spin diffusion theory and the spin conductance, the SMR ratio can be derived as
\begin{align}
\frac{\Delta \rho}{\rho} &= \theta_{\rm SH} \gamma \tanh (d/2\lambda), \\
\gamma &= \frac{4e^{2}}{\hbar} \frac{G_s}{S\sigma / \lambda},
\end{align}
where $e$ is the elementary charge $e>0$, $S$ is the surface area of the interface, $\lambda$ is the spin diffusion length, $d$ is the thickness of the NM, and $\gamma$ is the normalized spin conductance (we have assumed that $\gamma \ll 1$).
Thus, calculation of the SMR signal is attributed to that of the spin conductance $G_s$.

In the remaining part of Sec.~\ref{sec:formulation}, we formulate the spin current $I_s$ and the spin conductance $G_s$ as a function of the N\'{e}el vector of the AFI. 

\subsection{Normal metal}
\label{sec:NM}

We will describe a normal metal in terms of noninteracting electron system whose Hamiltonian is 
\begin{align}
H_{\rm{NM}}=\sum_{\bm{k}\sigma} \epsilon_{\bm{k}} c^{\dag}_{\bm{k}\sigma}c_{\bm{k}\sigma} .
\label{eq:Ham}
\end{align}
where $\epsilon_{\bm{k}}$ is kinetic energy \and $c_{{\bm k}\sigma}$ is the annihilation operator of conduction electrons with wavenumber ${\bm k}$ and spin ${\bm \sigma}$.
The spin accumulation at the interface is modeled by quasi-equilibrium distribution that is described by the effective Hamiltonian,
\begin{align}
\mathcal{H}_{\rm{NM}}&= H_{\rm NM} - \sum_{\sigma} \mu_\sigma N_{\sigma} ,
\end{align}
where $\mu_\sigma$ is the spin-dependent chemical potential and $N_\sigma$ is the number operator of conduction electrons with spin $\sigma \equiv  \pm 1$.
This effective Hamiltonian can be rewritten as
\begin{align}
\mathcal{H}_{\rm{NM}}=
\sum_{\bm{k}}(\xi_{\bm{k}} - \sigma
\delta \mu_s /2)c^{\dag}_{\bm{k}\sigma}c_{\bm{k}\sigma}.
\label{eq:calHam}
\end{align}
where $\xi_{\bm{k}}=\epsilon_{\bm{k}}-\mu$ is the kinetic energy measured from the averaged chemical potential $\mu=(\mu_\uparrow + \mu_\downarrow)/2$ and $\delta \mu = \mu_\uparrow - \mu_\downarrow$ is the spin chemical potential near the interface.

As shown in Sec.~\ref{sec:SpinCond2}, the spin conductance can be  written in terms of the local spin susceptibility, defined as
\begin{align}
\chi^{R}_{\rm{loc}}(\omega) &={\frac{1}{{N}_{\rm{NM}}}}\sum_{\bm{q}}
\int dt \, \chi^{R}_{+-}(\bm{q},t) e^{i\omega t},
\label{eq:3_spin_sus_loc} \\
    \chi^{R}_{+-}(\bm{q},t) &= {\frac{i}{{{N}_{\rm{NM}}} \hbar}}\theta(t)
    \langle [s^{+}_{\bm{q}}(t),s^{-}_{\bm{q}}(0)] \rangle , 
\end{align}
where $\langle \cdots \rangle = {\rm Tr}\, (e^{-\beta {\cal H}_{\rm NM}}\cdots)/{\rm Tr}\, e^{-\beta {\cal H}_{\rm NM}}$ indicates the average with respect to the quasi-equilibrium state, ${N}_{\rm{NM}}$ is the number of unit cells in normal metal, $s^{\pm}_{\bm{q}}$ are spin ladder operators defined as
\begin{align}
    s^{+}_{\bm{q}}&=(s^{-}_{\bm{q}})^{\dagger} = \sum_{\bm{k}} c^{\dag}_{\bm{k}+{\bm q}\uparrow}c_{\bm{k}\downarrow} ,
\end{align}
and $s^+_{\bm q}(t) = e^{iH_{\rm NM}t/\hbar} s^+_{\bm q}e^{-iH_{\rm NM}t/\hbar}$.
For the present model described by Eqs.~(\ref{eq:Ham}) and (\ref{eq:calHam}), the imaginary part of the local spin susceptibility is calculated as~\cite{Kato2020}
\begin{align}
    {\rm Im} \, \chi^{R}_{\rm{loc}}(\omega) 
    = \pi N(0)^2 (\hbar \omega + \delta \mu_s),
    \label{eq:chiloc}
\end{align}
where $N(0)$ is the density of states near the Fermi energy.

\subsection{Antiferromagnetic insulator}
\label{sec:AFI}

\begin{figure}[tb]
 \begin{center}
  \includegraphics[clip,width=5.0cm]{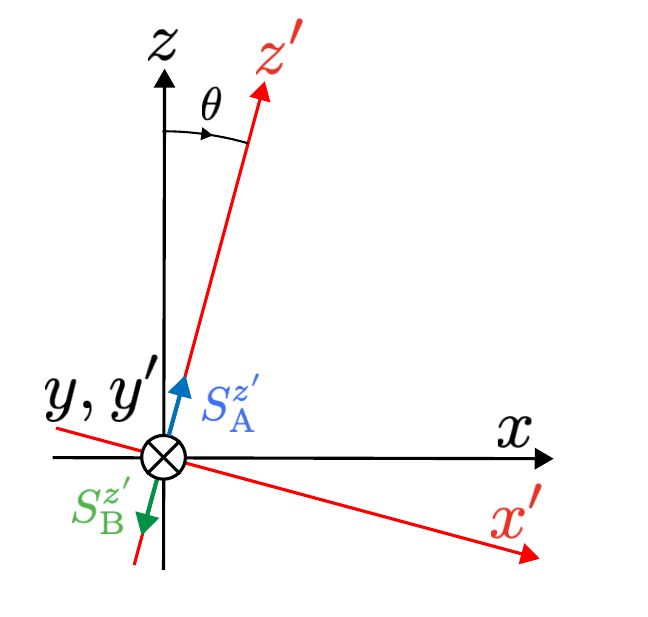}
    \caption{Coordinate transformation from the laboratory system $O$-$xyz$ to the coordinate system $O$-$x'y'z'$ with the $z'$-axis fixed to the magnetization orientation.}
    \label{fig:smr_axis}
  \end{center}
\end{figure}

Let us denote the azimuth angle of the N\'{e}el vector of the AFI measured from the $z$-axis as $\theta$.
We introduce a new magnetization-fixed coordinate systems $O$-$x'y'z'$, which is obtained by rotating the laboratory coordinates $O$-$xyz$ by $\theta$ around the $y$-axis as shown in Fig.~\ref{fig:smr_axis}.
Denoteing the components of the spin operator ${\bm S}$ in these two coordinate systems as $(S^x, S^y, S^z) $ and $(S^{x'}, S^{y'}, S^{z'})$, the transformation between them is expressed as
\begin{align}
\left( \begin{array}{c} S^{x'} \\ S^{y'} \\ S^{z'} \end{array}\right)
= \left(\begin{array}{ccc}
\cos \theta & 0 & - \sin \theta \\
0 & 1 & 0 \\
\sin \theta \, & 0 & \cos \theta \, \end{array} \right) 
\left( \begin{array}{c} S^{x} \\ S^{y} \\ S^{z} \end{array}\right)
.
\label{spin_coordinate_transformation}
\end{align}

For the AFI, we consider a quantum Heisenberg model in which localized spins are located on a cubic lattice.
We divide the lattice into A and B sublattices and define the spin operator at site $l$ on the sublattice $\nu$ ($=$A,B) as ${\bm S}_{\nu,l}$.
Accordingly, the Hamiltonian of the AFI can be described using the spin operators in the magnetization-fixed coordinates as
\begin{align}
H_{\rm AFI} = J \sum_{\langle {l,l'} \rangle}
\Biggl[
{\frac{1}{2}} \left( S^{+'}_{{\rm A},l}S^{-'}_{{\rm B},l'} +S^{-'}_{{\rm A},l}S^{+'}_{{\rm B},l'} \right) +S^{z'}_{{\rm A},l}S^{z'}_{{\rm B},l'} \Biggl] \nonumber \\
+ D \sum_{l} \Biggl[ (S^{z'}_{{\rm A},l})^2 + (S^{z'}_{{\rm B},l})^2 \Biggl]+ h \sum_{l} \Biggl[ S^{z'}_{{\rm A},l} - S^{z'}_{{\rm B},l} \Biggl],
\label{eq:HAFI}
\end{align}
where $\langle l,l'\rangle$ indicates a pair of nearest-neighbor sites, $S_{\nu,l}^{\pm'}=S_{\nu,l}^{x'}\pm i S_{\nu,l}^{y'}$ are spin ladder operators, and $J$ is the magnitude of the exchange interaction.
In order to fix the N\'{e}el vector in the quantum Monte Carlo simulation, we consider a small anisotropy in the magnetization and an alternating magnetic field, whose amplitudes are denoted by $D$ and $h$, respectively.
In the following discussion, we set $J$ as the unit of the energy.

The spin conductance is written in terms of local spin correlation functions defined as
\begin{align}
    G^{R,(a)}_{{\rm loc},\nu\nu'}(\omega) &= \frac{1}{N_{\rm AFI}} \sum_{\bm q} \int dt \, G^{R,(a)}_{\nu\nu'}(\bm{q},t) e^{i\omega t},
\end{align}
where $N_{\rm AFI}$ is the number of unit cells of the AFI and $G^{R,(a)}_{\nu\nu'}(\bm{q},t)$ ($a=1,2,3$) indicate three kinds of spin correlation functions, defined as
\begin{align}
G^{R,(1)}_{\nu\nu'}(\bm{q},t) &= -\frac{i}{\hbar} \theta(t)\langle{[S^{z'}_{\nu,\bm{q}}(t),S^{z'}_{\nu',{\bm{q}}}(0)]} \rangle,\\
G^{R,(2)}_{\nu\nu'}(\bm{q},t) &= -\frac{i}{\hbar} \theta(t)\langle{[S^{+'}_{\nu,\bm{q}}(t),S^{-'}_{\nu',\bm{q}}(0)]} \rangle, \\
G^{R,(3)}_{\nu\nu'}(\bm{q},t) &= -\frac{i}{\hbar}
    \theta(t)\langle{[S^{-'}_{\nu,\bm{q}}(t),S^{+'}_{\nu',\bm{q}}(0)]} \rangle.
\end{align}
Here, $\langle \cdots \rangle$ indicates the thermal average in the AFI, and 
the Fourier transformations of the spin operators are defined as
\begin{align}
S_{\nu,{\bm q}}^{+'} &= (S_{\nu,{\bm q}}^{-'})^\dagger =\sum_{l} S_{\nu,l}^{+'} e^{-i{\bm q}\cdot {\bm R}_{\nu,l}}, \\
S_{\nu,{\bm q}}^{z'} &= \sum_{l} S_{\nu,l}^{z'} e^{-i{\bm q}\cdot {\bm R}_{\nu,l}}, 
\end{align}
where ${\bm R}_{\nu,l}$ indicates the position of the site $l$ on the sublattice $\nu$.

\subsection{Microscopic description of spin conductance}
\label{sec:SpinCond2}

We consider an exchange coupling at the interface between the NM and AFI, whose Hamiltonian is given as
\begin{align}
    H_{\rm{ex}}=\sum_{\bm{k},\bm{q},\nu}[{\cal T}^\nu_{\bm{k},\bm{q}}{\mathcal{S}}^{+}_{\nu\bm{k}}s^{-}_{\bm{q}}+({\cal T}^\nu_{\bm{k},\bm{q}})^* {\mathcal{S}}^{-}_{\nu\bm{k}}s^{+}_{\bm{q}}],
 \label{eq:Hex}
\end{align}
using the laboratory coordinates, where ${\cal T}^\nu_{\bm{k},\bm{q}}$ is the magnitude of the exchange interaction at the interface.
We set ${\cal T}^\nu_{\bm{k},\bm{q}} = {\cal T}$ under the assumption that ${\cal T}^\nu_{\bm{k},\bm{q}}$ is independent of ${\bm k}$, ${\bm q}$, and $\nu$.
By using the transformation (\ref{spin_coordinate_transformation}),
we can rewrite the Hamiltonian in the magnetization-fixed coordinates as
\begin{align}
H_{\rm ex} &= \sum_{a=1}^3 H_{\rm{ex}}^{(a)}, 
\label{eq:Hex1} \\
    H_{\rm{ex}}^{(a)} &=g_{a}(\theta) \sum_{\bm{k},\bm{q},\nu}[{\cal T} {S}^{(a)}_{\nu\bm{k}}s^{-}_{\bm{q}}+{\cal T}^* ({S}^{(a)}_{\nu\bm{k}})^\dagger s^{+}_{\bm{q}}], \label{eq:Hex2}
\end{align}
where $S_{\nu{\bm k}}^{(a)}$ and $g_a(\theta)$ ($a=1,2,3$) are defined as
\begin{align}
    & {S}^{(1)}_{\nu\bm{k}}={S}^{z'}_{\nu\bm{k}},
    \quad g_{1}(\theta)=-\sin{\theta}, \label{eq:Hex3} \\
    & {S}^{(2)}_{\nu\bm{k}}={S}^{+'}_{\nu\bm{k}},
    \quad g_{2}(\theta)=\cos^2(\theta/2), \label{eq:Hex4} \\
    & {S}^{(3)}_{\nu\bm{k}}={S}^{-'}_{\nu\bm{k}},
    \quad g_{3}(\theta)=-\sin^2 (\theta/2). \label{eq:Hex5}
\end{align}
A detailed derivation is given in Appendix~\ref{app:SpinOperator}.

Performing a second-order perturbation with respect to $H_{\rm ex}$ yields the spin conductance as~\cite{Kato2020}
\begin{align}
G_s &= G_0 g_1(\theta)^2 \langle S_{\nu,l}^{z'} \rangle^2 \nonumber \\
&+\sum_{a=1}^{3} \sum_{\nu}
2 G_0 g_{a}(\theta)^2 \int \frac{d\epsilon}{2 \pi}\, \left. {\rm Im} \, \chi_{\rm loc}^R(\epsilon/\hbar) \right|_{\delta \mu_s=0}
    \nonumber \\ 
& \hspace{10mm} \times (-{\rm{Im}}\, G^{R,(a)}_{\nu\nu,\rm{loc}}(\epsilon/\hbar)) \left( - \frac{\partial f}{\partial \epsilon} \right),
\label{eq:Gs}
\end{align}
where $f(\epsilon)=(e^{\beta\epsilon}-1)^{-1}$ is the Bose distribution function and $G_0=2\pi |{\cal T}|^2 N_{\rm NM}^2 N_{\rm AFI} N(0)^2$ is a dimensionless parameter which represents the strength of the interfacial exchange coupling.
A detailed derivation is given in Appendix~\ref{app:SpinCurrent}.

The amplitude of the SMR signal is proportional to\footnote{In Ref.~\onlinecite{Hou2017}, the orientation of the N\'{e}el vector in the AFI is controlled by the magnetization of an adjacent YIG.
The negative SMR observed in this experiment can be reasonably understood if the N\'{e}el vector is perpendicular to the magnetization of YIG.
We also employ this assumption in our work.}
\begin{align}
    \Delta G &\equiv G(\theta = 0)-G(\theta = \pi/2).
\end{align}
For convenience of discussion, we will express $\Delta G$ by the sum of the two contributions, 
\begin{align}
& \Delta G = \Delta G_{z} + \Delta G_{xy}, 
\label{eq:Gsum} \\
& \Delta G_z/G_0 = - \langle S^{z'}_{\nu,l} \rangle^2 - 2J_1 , \\
& \Delta G_{xy}/G_0 = \frac{3}{2}J_2 -\frac{1}{2} J_3 , \\
& J_a = \sum_\nu \int_{-\infty}^\infty \frac{d\epsilon}{2\pi}
    \left[
    (-{\rm{Im}}\, G^{R,(a)}_{\rm{loc},\nu\nu}(\epsilon/\hbar))\frac{\beta \epsilon}{\sinh^2(\beta \epsilon/2)}
    \right], \label{eq:Jadef}
\end{align}
where $\Delta G_z$ and $\Delta G_{xy}$ are contributions from the spin-spin correlation in the $z$ and $xy$ directions, respectively.
We will show later that $\Delta G_z$ ($\Delta G_{xy}$) gives a negative (positive) contribution to the SMR signal. 
Note that the integrals, $J_a$ ($a=1,2,3$), vanish at zero temperature.
Therefore, a negative SMR signal $\Delta G/G_0 = -\langle S^{z'}_{\nu,l} \rangle^2$ is obtained at zero temperature consistently with the experiment~\cite{Hou2017}.
Thus, the SMR signal is formulated in terms of the spin correlation function $G_{\nu \nu,{\rm loc}}^{R,(a)}(\omega)$ and the staggered magnetization $|\langle S^{z'}_{\nu,l} \rangle|$.
In the next section, we explain a numerical method to calculate them.

\section{Numerical method}
\label{sec:Numerical}

We used the continuous-time quantum Monte Carlo (QMC) method to calculate the spin correlation functions and the staggered magnetization for a finite temperature~\cite{Evertz2003,Kawashima2004,Gubernatis2016}.
We used the program package, Discrete Space Quantum Systems Solver (DSQSS)~\cite{Motoyama2021,DSQSSWeb}, which implements a continuous-time path-integral QMC method based on a directed-loop algorithm~\cite{Syljuasen2002,Syljuasen2003} and the Suwa-Todo algorithm without the detailed balance condition~\cite{Suwa2010}.
To calculate the spin correlation functions $G_{\nu \nu,{\rm loc}}^{R,(a)}(\omega)$, we need to perform an analytic continuation from the Matsubara frequency $i\omega_n$ to the real frequency $\omega$.
However, the numerical analytic continuation is usually unstable and inaccurate.

\begin{figure}[tb]
 \begin{center}
    \includegraphics[clip,width=8.0cm]{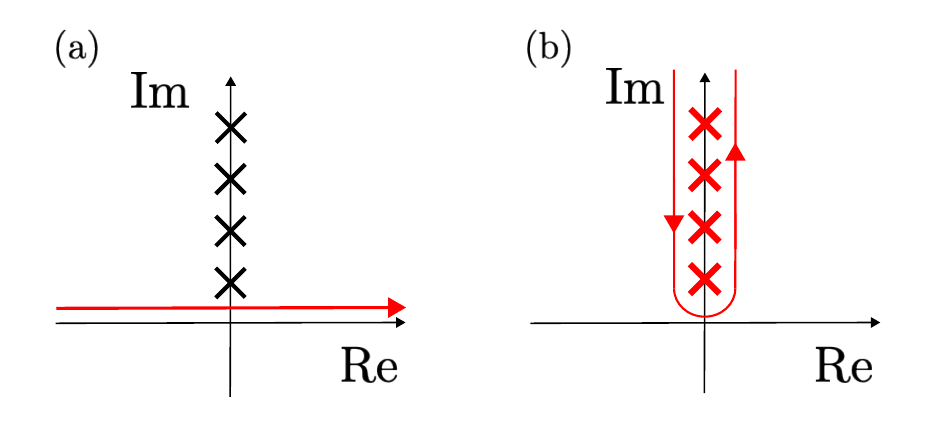}
    \caption{Change in integral path.
    (a) shows before the change to be integral path, where one needs to calculate all $G^{R}_{\rm{loc}}(\omega)$.  (b) is after the change to the path, where one calculates only  the Matsubara frequency $G^{R}_{\rm{loc}}(i\omega_{n})$ and $G^{\prime R,(a)}_{\rm{loc}}(i\omega_{n})$.}
    \label{fig:integral_change}
  \end{center}
\end{figure}

In this study, we employed a new method without performing a numerical analytic continuation and directly evaluated the integrals, $J_a$, defined by Eq.~(\ref{eq:Jadef}).
These integrals share a function $\beta \epsilon /\sinh^2(\beta \epsilon/2)$ that has double poles at $\epsilon = 2 i\pi n/\beta \equiv i\epsilon_n$, where $n$ is an integer
\footnote{$G^{R,(a)}_{\rm{loc}}(\epsilon/\hbar)$ vanishes at $\epsilon=0$ and therefore the integrand has no pole there.}.
Therefore, by modifying the integration path in the complex plane of $\epsilon$ as in Fig.~\ref{fig:integral_change}, $J_a$ can be expressed by the sum of the residues of the integrand at $\epsilon = i\epsilon_n$ ($n\ge 1$).
Since the residues are evaluated as
\begin{align}
    & R^{(a)}_{\nu,n} \equiv \underset{\epsilon = i\epsilon_{n}}{{\rm{Res}}}\frac{\beta \epsilon G^{R,(a)}_{\rm{loc},\nu\nu}(\epsilon/\hbar) e^{\beta  \epsilon}}{(e^{ \beta\epsilon}-1)^2}\nonumber \\
    & = \frac{1}{\beta} \left[ G^{R,(a)}_{\rm{loc},\nu\nu}(i\epsilon_{n}/\hbar) +  \epsilon_{n} \frac{dG_{\rm{loc},\nu\nu}^{R,(a)}(i\epsilon_n/\hbar)}{d\epsilon_n} \right] ,
\end{align}
we can express the integrals as

\begin{align}
    J_a = - \sum_{n=1}^\infty \sum_{\nu} {\rm Re}\, R^{(a)}_{\nu,n} .
    \label{eq:sumformula}
\end{align}
Thus, the integrals can be evaluated only from the information of the imaginary-times spin correlation functions.
The derivative of $G^{R,(a)}_{\rm{loc},\nu\nu}(i\epsilon_n/\hbar)$ with respect to the Matsubara frequency can be obtained by numerical differentiation using the Pade approximation.
Here, the Pade approximation is simply used for interpolation on the imaginary axis, and therefore, the accuracy of the simulation is greatly improved compared with direct numerical analytic continuation.

In our Monte Carlo simulation, we typically used $10^6$ Monte Carlo samples for each point.
In performing the sum in Eq.~(\ref{eq:sumformula}), we used numerical data on the spin correlation function for the Matsubara frequencies below a cutoff frequency and extrapolated it in the form $C/\epsilon_n^2$ ($C$: a constant) for higher Matsubara frequencies. 

\section{Results}
\label{sec:Result}

Here, we show the numerical results of the amplitude of the SMR signal, $\Delta G/G_0$, for $S=1/2$ and $S=1$ spin systems on a two-dimensional square lattice and a quasi-two-dimensional cubic lattice with a finite number of layers.
We also show the two contributions, $\Delta G_z/G_0$ and $\Delta G_{xy}/G_0$, separately (see Eqs.~\eqref{eq:Gsum}-\eqref{eq:Jadef}).
Finally, we discuss the effect of randomness by using a model with disordered exchange interactions.

We should note that the quasi-two-dimensional quantum Heisenberg model has no long-range order at finite temperatures because of the Mermin–Wagner theorem.
However, the staggered magnetization grows rapidly below a specific temperature, at which the correlation length exceeds the system size at low temperatures.
Since this rapid growth of the staggered magnetization is expected to simulate the actual behavior of thin antiferromagnet layers, our numerical results can be used for discussing the qualitative features of the SMR signal (for a detailed discussion, see Sec.~\ref{sec:discussion}).

\subsection{Case of $S=1/2$}
\label{sec:s1/2}

\begin{figure}[tb]
 \begin{center}
    \includegraphics[clip,width=8.0cm]{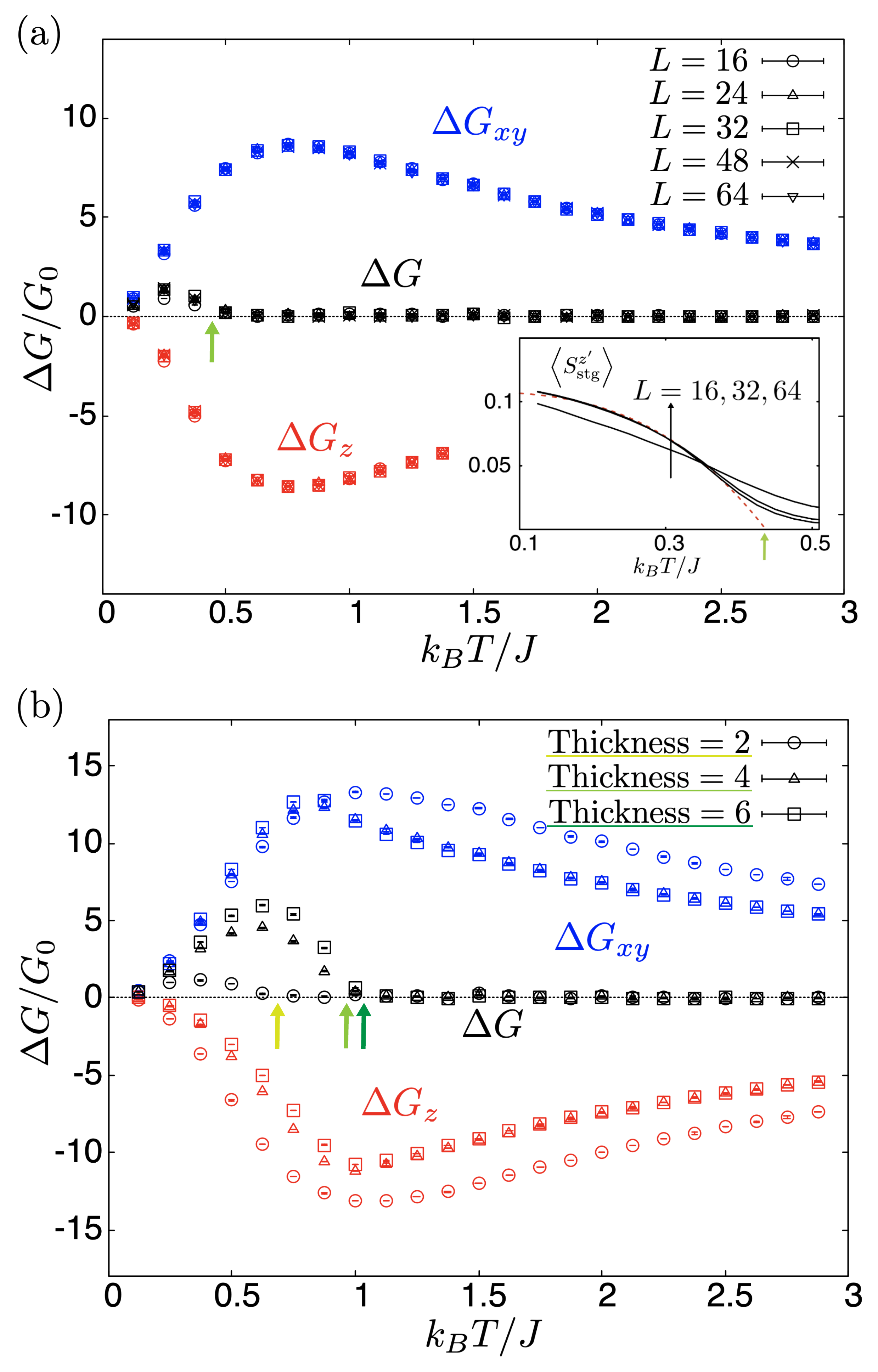}
    \caption{Temperature dependence of spin conductance $\Delta G/G_0 (= \Delta G_{xy}/G_0+\Delta G_z/G_0)$ for $S=1/2$ quantum Heisenberg model with
    alternating magnetic field $h=0.01J$ on (a) a $L\times L$ square lattice ($L=24,32,48,64$) and (b) a $16\times 16 \times W$ cubic lattice ($W=2,4,6$). The two contributions, $\Delta G_{xy}/G_0$ and $\Delta G_{z}/G_0$, are plotted separately.
    }
    \label{fig:smr_m1_2d_l}
  \end{center}
\end{figure}

First, we show the SMR signal for the $S = 1/2$ quantum Heisenberg model on a $L\times L$ square lattice ($L=24, 32, 48, 64$) in Fig.~\ref{fig:smr_m1_2d_l}~(a).
The black plots indicate the total SMR signal, $\Delta G/G_0$, whereas the red and blue plots indicate $\Delta G_z/G_0$ and $\Delta G_{xy}/G_0$, respectively.
To fix the N\'{e}el vector to the $z$ direction, we introduced a weak alternating magnetic field, $h=0.01J$ (see also Eq.~\eqref{eq:HAFI}).
We find that the size dependence of the SMR signal is weak for $L\ge 24$.
The solid curves in the inset of Fig.~\ref{fig:smr_m1_2d_l} are the staggered magnetization as a function of the temperature for $L=16,32,64$. 
We can see that the size dependence becomes weak for $L\ge 32$.
For a rough estimate of the temperature at which the staggered magnetization starts to grow, we performed a fitting of the form $|\langle S_i \rangle| \propto 1-(T/T_c)^\alpha$ in the range of
$|\langle S_i \rangle /S_0| > 0.2$ for $L=64$, as indicated by the red dashed line in the inset of Fig.~\ref{fig:smr_m1_2d_l}).
For convenience, this characteristic temperature is called the ordering temperature hereafter.
The estimated ordering temperature is indicated by the green arrow in the main graph of Fig.~\ref{fig:smr_m1_2d_l}.
Above the transition temperature, the SMR signal vanishes because $\Delta G_z/G_0$ and $\Delta G_{xy}/G_0$ cancel each other out.
This feature can be understood analytically from the high-temperature expansion (see Appendix~\ref{app:high-temp}).
In contrast, we find that the SMR signal becomes finite below the ordering temperature.
As the temperature is further lowered below the transition temperature, the SMR signal increases and then decreases toward zero temperature; the spin conductance has a peak roughly at three-fifths of the transition temperature.
Although the spin-wave approximation predicts the SMR signal to be negative at low temperatures~\cite{Kato2020}, such a sign change is not found at any point down to the lowest simulation temperature ($k_{\rm B}T/J=0.125$) in Fig.~\ref{fig:smr_m1_2d_l}~(a).

Figure~\ref{fig:smr_m1_2d_l}~(b) shows the results for a quasi-two-dimensional system with a finite thickness, i.e., a $16 \times 16 \times W$ cubic lattice ($W=2,4,6$).
As the thickness $W$ increases, the ordering temperature (indicated by the green arrows) increases and the spin conductance induced below the transition temperature becomes large.
We find that the peak of the spin conductance is greatly enhanced compared with that for a single-layer spin system ($W=1$).
Although the maximum of the spin conductance increases with increasing thickness $W$, the difference between the results for $W=4$ and $6$ is much smaller than that for $W=2$ and $4$.

\subsection{Case of $S=1$}
\label{sec:s1}

\begin{figure}[tb]
\begin{center}
\includegraphics[clip,width=8.0cm]{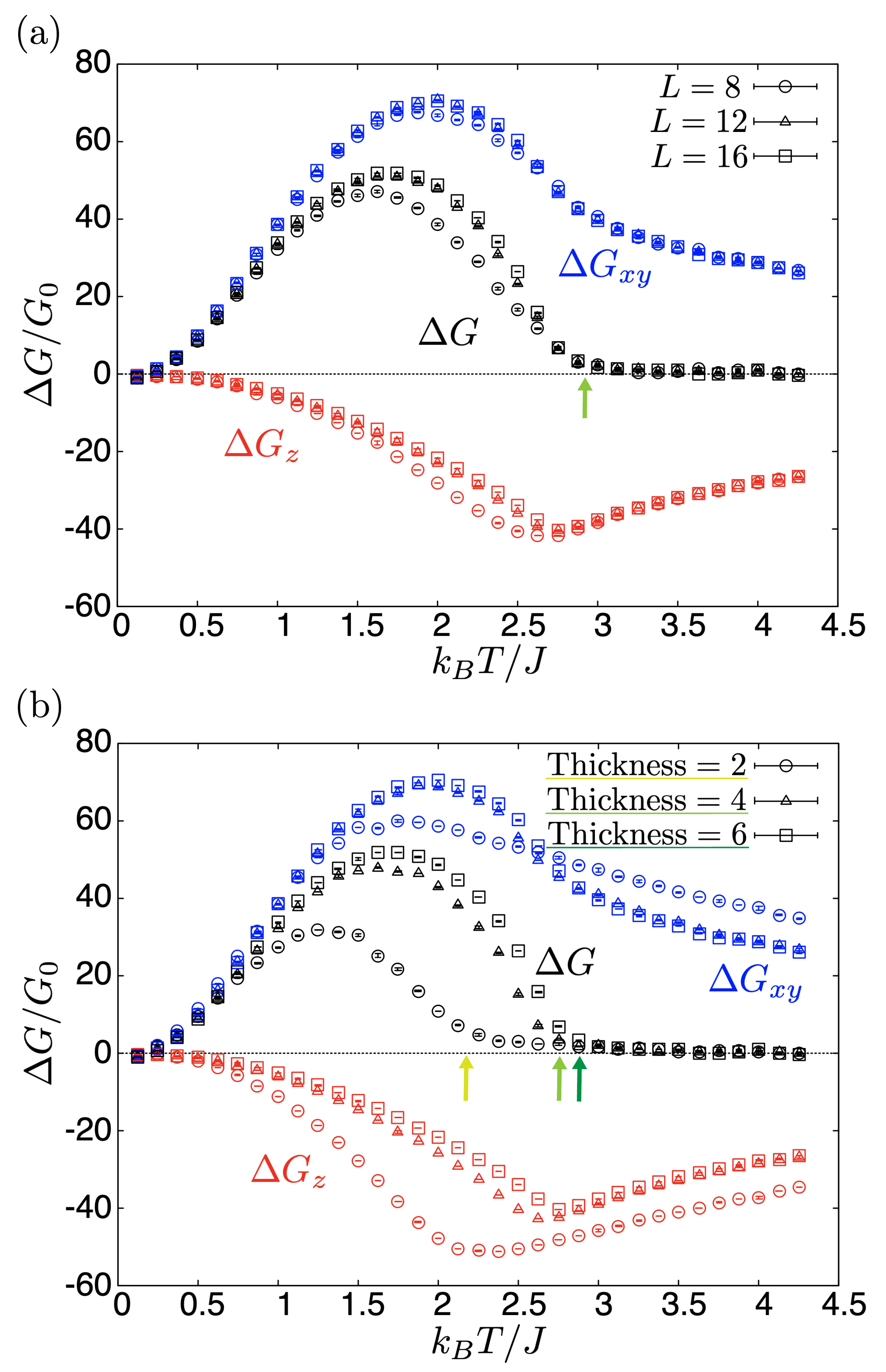}
\caption{Temperature dependence of spin conductance $\Delta G/G_0 (= \Delta G_{xy}/G_0+\Delta G_z/G_0)$ for $S=1$ quantum Heisenberg model with axial anisotropy $D=-0.1J$ on (a) a $L\times L\times 6$ cubic lattice ($L=8,12,16$) and (b) a $16\times 16 \times W$ cubic lattice ($W=2,4,6$). The two contributions, $\Delta G_{xy}/G_0$ and $\Delta G_{z}/G_0$, are plotted separately.}
\label{fig:smr_m2_3d_thi}
\end{center}
\end{figure}

Next, we consider the $S = 1$ quasi-two-dimensional quantum Heisenberg model with a finite thickness.
We introduce a small axial anisotropy $D=-0.1J$ for fixing the N\'{e}el vector.
The SMR signal for a $L\times L\times 6$ cubic lattice ($L=8, 12, 16$) is shown in Fig.~\ref{fig:smr_m2_3d_thi}~(a).
We find that the size dependence between $L=12$ and $L=16$ is weak enough for examining qualitative features.
The SMR signal for a $16\times 16\times W$ cubic lattice ($W=2, 4, 6$) is shown in Fig.~\ref{fig:smr_m2_3d_thi}~(b).
The ordering temperature (indicated by the green arrows) for the case of $S=1$ is higher than that of the case of $S=1/2$ (see Fig.~\ref{fig:smr_m1_2d_l}~(b)) because quantum fluctuations are suppressed.
As the thickness $W$ increases, the transition temperature raises and the SMR signal becomes large.
As in the case of $S=1/2$, the difference between the results for $W=4$ and $6$ is much smaller than that for $W=2$ and $4$.

Above the transition temperature, the SMR signal becomes small but remains finite in contrast to the case of $S=1/2$.
This feature can be understood by the high-temperature expansion (see Appendix~\ref{app:high-temp}).
As the temperature decreases below the ordering temperature, the SMR signal increases, takes a maximum roughly at three-fifths of the ordering temperature, and then decreases.
The maximum value of the signal is rather larger than in the case of $S=1/2$.
At sufficiently low temperatures, the SMR signal becomes negative, as predicted by the spin-wave approximation (for a detailed comparison, see Appendix~\ref{app:sw}).

\subsection{Disordered case for $S=1$}
\label{sec:disordered}

Finally, let us consider the effect of randomness by introducing disordered exchange interactions.
We consider disordered exchange interactions whose probability density function~\cite{Laflorencie_random_2006} is given by
\begin{align}
P(J) =  J^{-1+\delta^{-1}} \delta^{-1} \Theta (J) \Theta (1-J),
\label{eq6:dis}
\end{align}
where $\Theta(x)$ is the Heaviside step function and $\delta$ is a parameter
which we set to 1.5 in our simulation.

\begin{figure}[tb]
 \begin{center}
    \includegraphics[clip,width=8.0cm]{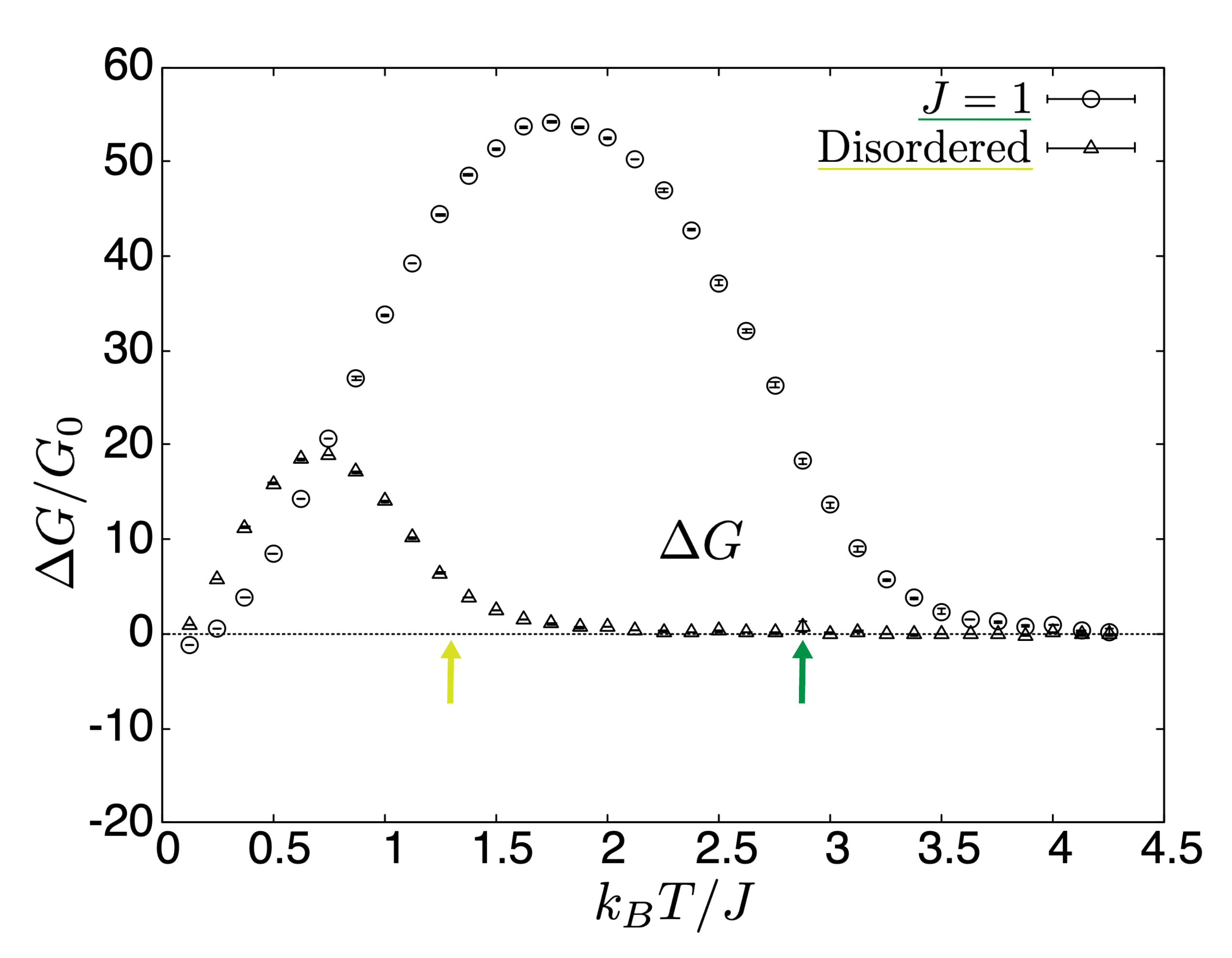}
\caption{Temperature dependence of $\Delta G/G_0$ for disordered $S=1$ quantum Heisenberg model on a $16\times 16 \times 6$ cubic lattice.
The parameter $\delta$ in the probability distribution (\ref{eq6:dis}) was set at $1.5$.}
    \label{fig:smr_m2_3d_dis}
  \end{center}
\end{figure}

Figure~\ref{fig:smr_m2_3d_dis} shows the SMR signal for the disordered $S=1$ quantum Heisenberg model on a $16\times 16 \times 6$ cubic lattice.
The two legends indicate the results for the uniform and disordered cases, respectively.
Note that the value of the exchange interaction $J$ is taken to be a constant in the uniform system.
In the disordered case, the ordering temperature is suppressed and the SMR signal becomes small.
However, the qualitative features are common to the uniform case; the SMR signal becomes small above the transition temperature, whereas it increases and then decreases as the temperature falls below the ordering temperature.

\section{Experimental Relevance}
\label{sec:discussion}

Here, we discuss how our simulation is related to the SMR measurements.
As already stated, our calculation has several limitations in comparing it with experimental results.
Because of the Mermin-Wagner theorem~\cite{Mermin66}, the quasi-two-dimensional quantum spin systems never have a finite-temperature phase transition, whereas its correlation length diverges toward zero temperature.
To capture this feature of quasi-two-dimensional quantum systems, we would need to perform a  large-scale QMC simulation on a large system and to carefully analyze numerical data by using finite-size scaling.
Furthermore, we incorporated additional parameters, i.e., a staggered magnetic field and an anisotropy, in our model to fix the direction of the AFI N\'{e}el vector.
These additional terms in the Hamiltonian, as well as finite-size effect, smear out the rapid rise of the staggered magnetization near the ordering temperature.
On the other hand, the AFI/NM bilayer used in the experiments has, more or less, a finite size.
When using experimental methods such as a neutron scattering, we usually observe that the staggered magnetization of the AFI layer grows rapidly below a specific finite temperature, which is often called the transition temperature.
In this sense, our simulation, in which the antiferromagnetic ordering grows below a specific temperature due to the finite-size effect, resembles the behavior in real experiments.
Therefore, the qualitative features of our simulation are expected to be observed experimentally.

The characteristic features found in our simulations are summarized as follows.
(a) When the magnitude of the localized spin, $S$, decreases, the SMR signal becomes small.
In particular, the SMR signal is rather suppressed for the $S=1/2$ case.
(b) The SMR signal becomes large below the transition temperature.
The peak of the SMR signal {\it does not} correspond to the transition temperature.
(c) The SMR signal becomes negative at sufficiently low temperatures for the $S=1$ case.
This feature is consistent with the previous theory based on the spin-wave approximation~\cite{Kato2020}.
Although we expect that the sign change of SMR also occurs at low temperatures for the  $S=1/2$ case, we need to perform QMC simulation at lower temperatures to check it.
(d) Although the randomness of exchange interactions reduces both the transition ordering temperature and the SMR signal, it does not change the qualitative features of the temperature dependence of the SMR.

The features obtained in our simulation are consistent with an experiment on a NiO thin layer~\cite{Hou2017}, in which there was a $S=1$ spin at each Ni site.
However, there are also differences; the SMR never becomes zero even at high temperatures and the curves of the SMR signal in our simulation do not completely fit to the experimental results.
We attribute these differences to
additional factors, such as domain formation, roughness of interfaces, the proximity effect due to adjacent magnetic materials, and defects in real AFI layers.
A more realistic simulation that considers these factors is left as a future study.

\section{Summary}
\label{sec:summary}

We studied the temperature dependence of the amplitude of SMR by using a quantum Monte Carlo simulation.
Following Ref.~\onlinecite{Kato2020},
we formulated the spin conductance at the AFI/NM interface in terms of the retarded component of the local spin susceptibility and calculated it by using an improved quantum Monte Carlo method  that does not need a direct numerical analytic continuation.
We showed that the SMR starts to grow below a specific temperature at which the correlation length reaches the system size.
As the temperature is lowered, the SMR increases, takes a maximum value roughly at three-fifths of the ordering temperature, and then decreases.
We discussed the dependence of the magnitude of the spin and thickness of the AFI layer as well as the effect of the disordered exchange interactions.
These qualitative features are expected to be observed experimentally.

Our simulation using a simplified model is expected to be a useful starting point for understanding SMR in quasi-two-dimensional quantum spin systems.
A detailed comparison with experiments using more realistic models, as well as a detailed finite-size analysis with a large-scale QMC simulation, will be left for a future study.

\section*{Acknowledgements}
T. K. acknowledges support from the Japan Society for the Promotion of Science (JSPS KAKENHI Grants No. 20K03831). M. M. is financially supported by a Grant-in-Aid for Scientific Research B (20H01863) from MEXT, Japan. T. I. is supported by International Graduate Program of Innovation for Intelligent World (IIW) of The Univ. of Tokyo.
T. K. and T. I. acknowledge K. Yoshimi and Y. Motoyama for giving us useful information on the detailed implementation of QMC in the DSQSS package.

\appendix

\section{Rotation of spin operators}
\label{app:SpinOperator}

We define the spin ladder operators in the laboratory coordinate as $S_{\nu,l}^{\pm}=S_{\nu,l}^{x}\pm i S_{\nu,l}^{y}$ and their Fourier transformations as
\begin{align}
S_{\nu,{\bm q}}^{+} &= (S_{\nu,{\bm q}}^{-})^\dagger =\sum_{l} S_{\nu,l}^{+} e^{-i{\bm q}\cdot {\bm R}_{\nu,l}},
\end{align}
where ${\bm R}_{\nu,l}$ is the position of site $l$ on the sublattice $\nu$.
Using Eq.~(\ref{spin_coordinate_transformation}), these Fourier transformations are related to those in the magnetization-fixed coordinate as 
\begin{align}
S^{+}_{\nu{\bm k}}=\cos^{2}(\theta/2) S^{+'}_{\nu{\bm k}}-
\sin^{2}(\theta/2) S^{-'}_{\nu{\bm k}}- \sin \theta \, S^{z'}_{\nu{\bm k}},\\
S^{-}_{\nu{\bm k}}=\cos^{2}(\theta/2) S^{-'}_{\nu{\bm k}}-\sin^{2}(\theta/2) S^{+'}_{\nu{\bm k}}-\sin \theta \, S^{z'}_{\nu{\bm k}}.
\end{align}
By substituting these equations into Eq.~(\ref{eq:Hex}), we obtain Eqs.~(\ref{eq:Hex1})-(\ref{eq:Hex5}).

\section{Spin Current and Spin Conductance}
\label{app:SpinCurrent}

Here, we derive an analytic expression for spin conductance following Ref.~\onlinecite{Kato2020}.
The spin current operator $\hat{I}_{S}$ is defined by the spin loss rate in the NM:
\begin{align}
    \hat{I}_{S}&=-\hbar\partial_{t}s^{z}_{\rm{tot}}=i[s^{z}_{\rm{tot}},H_{\rm{ex}}],\\
    s^{z}_{\rm{tot}}&=\frac{1}{2}\sum_{\bm{k}}(c^{\dag}_{\bm{k}\uparrow}c_{\bm{k}\uparrow} + {\rm h.c.} )
\end{align}
Using Eqs.~(\ref{eq:Hex})-(\ref{eq:Hex}), the current operator is rewritten as
\begin{align}
    \hat{I}_{S}&=\sum_{a=1}^3 \hat{I}_{S}^{(a)} ,\\
    \hat{I}_{S}^{(a)}&= i[s^{z}_{\rm{tot}},H_{\rm{ex}}^{(a)}] \nonumber \\
    &=-i g_{a}(\theta) \sum_{\bm{k},\bm{q},\nu}[T^\nu_{\bm{k},\bm{q}}{S}^{(a)}_{\bm{k}}s^{-}_{\bm{q}}-{\rm h.c.}] .
\end{align}
The spin current operator is expressed by a formal series of the perturbative Hamiltonian $H_{\rm ex}$ as
\begin{align}
    \langle{\hat{I}_{S}^{(a)}}\rangle 
    &={\rm{Re}}
    \Biggl[
    -2i g_{a}(\theta) \sum_{\bm{k},\bm{q}}
    T_{\bm{k},\bm{q}} 
    \Bigl\langle T_{C} {S}^{(a)}_{\nu \bm{k}}(\tau_1)s^{-}_{\bm{q}}(\tau_2) \nonumber \\
    &\hspace{10mm} \times {\rm{exp}}
    \left(-
    \frac{i}{\hbar}\int_{C}d\tau H_{\rm{ex}}^{(a)}(\tau)
    \right)\Bigr\rangle
    \Biggl] .
\end{align}
By expanding the exponential function and by taking the terms up to the first order of $H_{\rm ex}^{(a)}$, the spin current is calculated as
\begin{align}
\langle{\hat{I}_{S}^{(a)}}\rangle 
    &=-2\hbar N_{\rm NM} g_a(\theta)^2 \int \frac{d\epsilon}{2\pi} \, {\rm{Re}} \, \sum_{{\bm k},{\bm q},\nu} 
    |{\cal T}_
    {{\bm k},{\bm q}}|^2  \nonumber \\
    &\times \biggl[ \chi^<({\bm q},\epsilon/\hbar) G^{R,(a)}_{\nu\nu} ({\bm k},\epsilon/\hbar) 
    \nonumber \\
    & \hspace{5mm} - \chi^A({\bm q},\epsilon/\hbar) G^{<,(a)}_{\nu\nu} ({\bm k},\epsilon/\hbar) \biggr],
\end{align}
where the superscripts, $<$ and $A$, indicate the lesser and advanced components of the correlation functions.
Using the fluctuation dissipation theorems,
\begin{align}
    \chi^{<}({\bm{q}},\epsilon/\hbar) &= 2 i f(\epsilon+\delta\mu_s)\, {\rm{Im}}\, \chi^{R}({\bm{q}},\epsilon/\hbar),\\
    G^{<,(a)}_{\nu\nu'}({\bm{k}},\epsilon/\hbar) &= 
    - 2\pi i N_{\rm AFI}\langle S_{\nu,l} \rangle^2 
    \delta_{a,1} \delta_{{\bm k},{\bm 0}} \nonumber \\
    &+ 2 i f(\epsilon)\, {\rm{Im}}\, G^{R,(a)}_{\nu\nu'}({\bm{k}},\omega),
    \label{eq:FDTheorem2}
\end{align}
and $\chi^A({\bm q},\epsilon/\hbar) = (\chi^R({\bm q},\epsilon/\hbar))^*$, we obtain
\begin{align}
    \langle {\hat{I}^{(a)}_{S}} \rangle &= 
    I_{S,1} + I_{S,2}, \\
    I_{S,1} &= \hbar A \sin^2 \theta \langle S_{\nu,l}^{z'} \rangle^2 \, {\rm Im} \, \chi_{\rm loc}^R(0) ,\\
    I_{S,2} &= \sum_{a=1}^3 \sum_{\bm{k},\bm{q},\nu} 2 \hbar A g_{a}(\theta)^2 \int \frac{d\epsilon}{2\pi} \, {\rm{Im}}\, \chi_{\rm loc}^{R}(\bm{q},\epsilon/\hbar) \nonumber \\
    & \times (-\, {\rm{Im}}\, G_{{\rm loc},\nu\nu}^{R,(a)}(\bm{k},\epsilon/\hbar))[f(\epsilon)-f(\epsilon+\delta\mu_{s})],
    \label{eq:spincurrentformula}
\end{align}
where $A=4|{\cal T}|^2 N_{\rm NM}^2 N_{\rm AFI}$ and $f(\epsilon)=(e^{\beta \epsilon}-1)^{-1}$ is the Bose distribution function.
Using the definition of the spin conductance, Eq.~(\ref{eq:spinconductancedef}), we obtain Eq.~(\ref{eq:Gs}).

\section{High-temperature limit}
\label{app:high-temp}

Here, we derive analytic results by considering the leading term in the high-temperature region.
At sufficiently high temperatures, the system Hamiltonian of the AFI can be approximated by a one-site Hamiltonian because the correlation length becomes very short.
Assuming zero staggered magnetization, the approximate Hamiltonian is
\begin{align}
H_{\rm AFI} &= D \sum_{l} \Biggl[ (S^{z'}_{{\rm A},l})^2 + (S^{z'}_{{\rm B},l})^2 \Biggl]\nonumber \\
&+ h \sum_{l} \Biggl[ S^{z'}_{{\rm A},l} - S^{z'}_{{\rm B},l} \Biggl].
\label{eq:mf_HAFI}
\end{align}
Denoting the eigenstates and eigenenergies of this Hamiltonian with $\ket{n}$ and 
$E_n$, respectively, the Lehmann representation of spin susceptibility can be expressed as
\begin{align}
G^{R,(1)}(\bm{q},\omega) 
={\frac{-i}{ Z \hbar}}
    \sum_{n,m}
    \frac{
    \left(e^{-\beta E_n} - e^{-\beta E_m} \right)
    | \left< n \right| S^{z'}_{\bm{q}} \left| m \right> |^{2}}
    {\hbar \omega+E_{n}-E_{m}+i \delta}.
\end{align}
where $\delta$ is a positive infinitesimal.
At high temperatures ($\beta \gg 1$), the leading term of the SMR is given as
\begin{align}
    & \Delta G_z/G_0 \nonumber \\
    &= \frac{2\pi}{Z}
    \sum_{n,m}\left| \braket{ n |
    S^{z'}_{\bm{q}}
    | m }
    \right|^{2} \nonumber \\
    &- \frac{\pi}{Z}\sum_{n,m}
    \left\{ \beta (E_{m}+ E_{n}) \right\}
    \left| \braket{ n | S^{z'}_{\bm{q}} | m } \right|^{2}.
\end{align}
By direct calculation of the Hamiltonian (\ref{eq:mf_HAFI}), we obtain the analytic form of the SMR at high temperatures as
\begin{align}
\Delta G_z/G_0
&= \left\{ \begin{array}{ll} 0, & (S=1/2), \\
\displaystyle{\frac{-4 \pi \beta D}{1+2e^{\beta D}}}, & (S=1). \end{array} \right.
\end{align}
For $S=1$, we obtain $\Delta G_z/G_0 = 0.11$ for $D=-0.1J$ and $k_B T/J = 4$.

\section{Spin-wave approximation}
\label{app:sw}

\begin{figure}[tb]
\begin{center}
\includegraphics[clip,width=8.0cm]{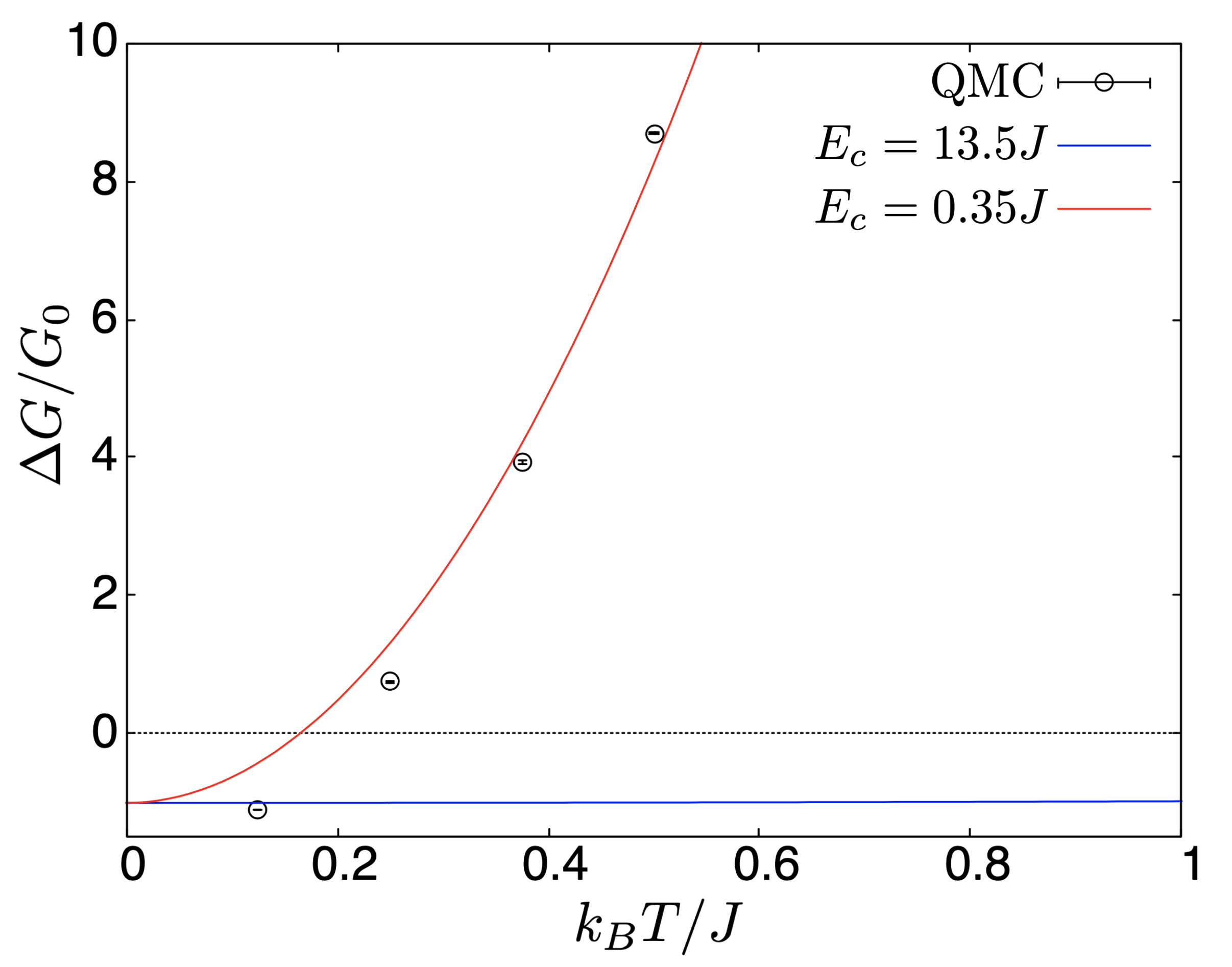}
\caption{Comparison of QMC simulation and spin-wave approximation for the $S=1$ quantum Heisenberg model on a $16\times 16 \times 6$ cubic lattice.
The blue curve indicates the prediction of the spin-wave theory given by Eq.~(\ref{eq:spinw}) with a cut-off energy of $E_{\rm c}=13.5J$.
The red curve indicates the one modified by replacing the cut-off energy with $E_{\rm c}=0.35J$.}
\label{fig:appendix_spinwave}
\end{center}
\end{figure}

In Ref.~\onlinecite{Kato2020}, the SMR is calculated within the spin-wave approximation as
\begin{align}
    \frac{\Delta G_z}{G_0}  =  -1+\frac{4.4}{S}
    \left( \frac{k_B T}{E_c} \right)^2,
    \label{eq:spinw}
\end{align}
where $E_c$ is the cut-off energy and $S$ is the amplitude of the spin. 
Within the spin-wave approximation, the cut-off is given by $E_c=\hbar v_m k_c=\hbar \cdot 2\sqrt{3} J S_0 a/\hbar \cdot (6 \pi)^{1/3}/a \simeq 13.5J$.
Figure~\ref{fig:appendix_spinwave} show the QMC data and the prediction of the spin-wave theory for the $S=1$ case.
If we employ $E_c= 13.5J$, the spin-wave theory (indicated by the blue curve in Fig.~\ref{fig:appendix_spinwave}) does not fit the QMC data.
This disagreement is expected to be due to the smallness of the spin amplitude ($S=1$);
strong quantum fluctuations will modify the effective amplitude of the spin, which is represented by $S$ in Eq.~(\ref{eq:spinw}). 
This effect can be taken into account by modifying the cut-off energy.
If we modify the cut-off energy to $E_c= 0.35J$, the spin-wave theory becomes consistent with the QMC data, as indicated by the red curve.
Noting that the QMC data include the finite-size effect in the temperature range of Fig.~\ref{fig:appendix_spinwave}, the agreement between the red curve and the QMC data is satisfactory.

\bibliography{ref}

\end{document}